\documentclass[10pt]{article} 

\usepackage[preprint]{tmlr}
\PassOptionsToPackage{numbers, compress}{natbib}
\usepackage{graphicx}
\usepackage{multicol,multirow}
\usepackage{mathrsfs}
\usepackage{amsthm}
\usepackage{rotating}
\usepackage{appendix}
\usepackage{nomencl}
\usepackage{etoolbox}

\usepackage[numbers]{natbib}
\usepackage{xcolor}
\usepackage{lipsum}
\usepackage{placeins}

\usepackage[utf8]{inputenc} 
\usepackage[T1]{fontenc}    
\usepackage{url}            
\usepackage{booktabs}       
\usepackage{amsfonts}       
\usepackage{nicefrac}       
\usepackage{microtype}      
\usepackage{xcolor}         

\usepackage{bm}             
\usepackage{amsmath}
\usepackage{amssymb}
\usepackage{cprotect}
\usepackage{dblfloatfix}
\usepackage{subcaption}
\usepackage{float}

\numberwithin{equation}{section}

\newcommand{\blank}{\mathrel{\;\cdot\;}}

\DeclareMathOperator*{\argmin}{argmin}

\newcommand\corestr[2]{{
  \left.\kern-\nulldelimiterspace 
  #1
  \vphantom{\big|}
  \right|^{#2}
}}

\title{Physics-constrained convolutional neural networks for inverse problems in spatiotemporal partial differential equations}

\author{\name Daniel Kelshaw
		\email djk21@ic.ac.uk\\
		\addr Department of Aeronautics, Imperial College London, London, UK 
		\AND
		\name Luca Magri
		\email l.magri@imperial.ac.uk\\
		\addr Department of Aeronatics, Imperial College London, London, UK\\ The Alan Turing Institute, British Library, London, UK \\ Politecnico di Torino, DIMEAS, Torino, Italy}

\def\month{08}  
\def\year{2024} 

\begin{document}

\maketitle

\begin{abstract}
We propose a physics-constrained convolutional neural network (PC-CNN) to solve two types of inverse problems in partial differential equations (PDEs), which are nonlinear and vary both in space and time. In the first inverse problem, we are given data that is offset by spatially varying systematic error (i.e., the bias, also known as the epistemic uncertainty). The task is to uncover the true state, which is the solution of the PDE, from the biased data. In the second inverse problem, we are given sparse information on the solution of a PDE. The task is to reconstruct the solution in space with high-resolution. First, we present the PC-CNN, which constrains the PDE with a time-windowing scheme to handle sequential data. Second, we analyse the performance of the PC-CNN for uncovering solutions from biased data. We analyse both linear and nonlinear convection-diffusion equations, and the Navier-Stokes equations, which govern the spatiotemporally chaotic dynamics of turbulent flows. We find that the PC-CNN correctly recovers the true solution for a variety of biases, which are parameterised as non-convex functions. Third, we analyse the performance of the PC-CNN for reconstructing solutions from sparse information for the turbulent flow. We reconstruct the spatiotemporal chaotic solution on a high-resolution grid from only $\lesssim 1\%$ of the information contained in it. For both tasks, we further analyse the Navier-Stokes solutions. We find that the inferred solutions have a physical spectral energy content, whereas traditional methods, such as interpolation, do not. This work opens opportunities for solving inverse problems with partial differential equations. 
\end{abstract}

\section{Introduction}

Physical and engineering phenomena can be modelled by nonlinear partial differential equations, which may vary both in space and time. In partial differential equations, on the one hand, the goal of the \textit{forward problem} is to predict the output (for example, the evolution of the system) given inputs on the initial conditions, system parameters, and boundary conditions. On the other hand, the goal of the \textit{inverse problem} is to compute the input that corresponds to a given set of outputs. Whereas forward problems are typically well-posed, inverse problems are not. The overarching objective of this paper is to propose a method --- the physics-constrained convolutional neural networks (PC-CNN) --- to solve two types of inverse problems of engineering interest: \textit{(i)} uncovering the solution to the partial differential equation that models the system under investigation from biased data; and \textit{(ii)} the reconstruction of the solution on a high-resolution grid from sparse information.

The first inverse problem concerns error removal, which, in the literature, typically considers only the case of aleatoric noise-removal, which is a term attributed to methods that remove small, unbiased stochastic variations from the underlying solution. This has been achieved through various methods, including filtering methods \citep{Kumar2017}; proper orthogonal decomposition \citep{Raiola2015, Mendez2017}; and the use of autoencoders \citep{vincent2010StackedDenoisingAutoencoders}, to name only a few. Our first demonstration of the proposed physics-constrained convolutional neural network aims, instead, to formalise and solve an inverse problem to remove large-amplitude bias (systematic error) from data; ultimately, uncovering the true solution of the partial differential equation.

The second inverse problem concerns the reconstruction of solutions to partial differential equations from sparse information, which describes only a small portion of the solution and is a challenge for system identification~\citep[e.g.,][]{Brunton2016}. Current methods primarily make use of convolutional neural networks due to their ability to exploit spatial correlations. The current data-driven approaches need access to the full high-resolution samples in the training, which are needed to produce a parametric mapping for the reconstruction task~\citep[e.g.,][]{Dong2014, Shi2016, Yang2019, Liu2020}. In many cases, access to the high-resolution samples is limited, on which the method proposed in this paper does not rely.

Common to both inverse problems that we tackle in this paper is the goal of finding a mapping from some known quantity to the corresponding solution of the governing partial differential equation. By exploiting the universal function approximation property of neural networks, it is possible to obtain numerical surrogates for function mappings or operators~\citep{hornik1989MultilayerFeedforwardNetworks, Zhou2020}. For a network parameterised by weights, there exists an optimal set of weights that results in the desired mapping; the challenge being to realise these through an optimisation process. This optimisation is formulated as the minimisation of a loss function, the evaluation of which offers a distance metric to the desired solution. When considering forms of structured data, convolutional neural networks excel due to their ability to exploit spatial correlations in the data, which arise from the use of localised kernel operations.

With physical systems governed by partial differential equations, the problems encountered in the absence of ground-truth high-resolution observations can be mitigated by employing a physics-constrained approach by imposing prior knowledge of the governing equations~\citep[e.g.,][]{Lagaris1998,raissi2019PhysicsinformedNeuralNetworks,doan2021ShortandLongtermPredictions}. One such tool is that of Physics-informed neural networks (PINNs)~\citep{raissi2019PhysicsinformedNeuralNetworks}, which provide a tool for physically-motivated problems, exploiting automatic-differentiation to constrain the governing equations. Although the use of PINNs for inverse problems shows promising results for simple systems~\citep{eivazi2022PhysicsinformedDeeplearningApplications}, they remain challenging to train, and are not designed to exploit spatial correlations~\citep{krishnapriyan2021characterizing, grossmann2023can}. On the other hand, convolutional neural networks are designed to exploit spatial correlations, but they cannot naturally leverage automatic differentiation to evaluate the physical loss, as PINNs do, because they provide a mapping between states, rather than a mapping from the spatiotemporal coordinates to states as in PINNs. As such, the design of physics-informed convolutional neural networks is more challenging, and relies on finite-different approximations or differentiable solvers~\citep{kelshaw2023UncoveringSolutionsData, Gao2021}. For example, the authors of \citep{Gao2021} show results for a steady flow field (with no temporal dynamics), which produces a mapping for stationary solutions of the Navier-Stokes equations.

In this work, we propose a physics-constrained convolutional neural network with time-windowing suitable for handling a range of physically-motivated inverse problems. Imposing prior knowledge of the physics allows us to restrict the function space and use limited collocation points to find a solution. We emphasise that this is not a PINN approach~\cite{raissi2019PhysicsinformedNeuralNetworks} because we do not leverage automatic differentiation to obtain gradients of outputs with respect to inputs to constrain the physics. This network employs a time-windowing scheme to effectively compute the time-dependent residuals without resorting to recurrent-based network architectures. Realisations of the network that do not conform to the given governing equations are penalised, which ensures that the network learns to produce physical predictions. 

We first provide the necessary background in Sec.~{\ref{sec:background}}, and introduce the two physically-motivated inverse problems. A brief background on convolutional neural networks is provided in Sec.~\ref{sec:cnn_overview}, before introducing the proposed physics-constrained convolutional neural networks in Sec.~\ref{sec:pc_cnn_overview}. Here, we outline the generic approach before discussing specifics for each of the inverse problems discussed in the paper. In Sec.~\ref{sec:spectral_method} we discuss data generation, and how employing a pseudospectral solver can be used to further constrain physical properties of the system. Finally, we showcase results for each method. Results for the bias-removal task are shown in Sec.~\ref{sec:results:bias_removal}, where the method is demonstrated on three partial differential equations of increasing complexity: linear convection-diffusion, nonlinear convection-diffusion, and the Navier Stokes equations. Results for the reconstruction task are shown in Sec.~\ref{sec:results:reconstruction}, in which results are analysed on the turbulent chaotic Navier Stokes equations. Conclusions in Sec.~\ref{sec:conclusion} mark the end of the paper.

\section{Background} \label{sec:background}

We consider dynamical systems, which are governed by partial differential equations (PDEs) of the form
\begin{equation} \label{eqn:dynamical_system}
    \mathcal{R}({\bm {\tilde u}}; {\bm p}) \equiv \partial_t {\bm {\tilde u}({\bm x},t)} - \mathcal{N}({\bm {\tilde u}}; {\bm p}),
\end{equation}
where ${\bm x} \in \Omega \subset \mathbb{R}^n$ is the spatial location;  $n$ is the space dimension; and $\Omega$ is the spatial domain. Time is denoted by $t \in [0, T] \subset \mathbb{R}_{\geq 0}$; ${\bm {\tilde u}}({\bm x},t): \Omega \times [0, T] \rightarrow \mathbb{R}^{m}$ is the state, where $m$ is the number of components; and ${\bm p}$ is a vector that encapsulates the physical parameters of the system. A sufficiently smooth differential operator is denoted by $\mathcal{N}$, which represents the nonlinear part of the system of $m$ partial differential equations; $\mathcal{R}$ is the residual; and $\partial_t$ is the partial derivative with respect to time.
A solution ${\bm u}$ of the PDE~\eqref{eqn:dynamical_system} is the state that makes the residual vanish, i.e. ${\bm u}$ is such that $\mathcal{R}({\bm u}; {\bm p}) = 0$, with Eq.~\eqref{eqn:dynamical_system} being subject to prescribed boundary conditions on the domain boundary, $\partial\Omega$, and initial conditions at $t=0$. 

\subsection{Uncovering solutions from biased data} \label{sec:bias_removal_methodology}

\begin{figure*}[ht]
    \centering
    \includegraphics[width=\linewidth]{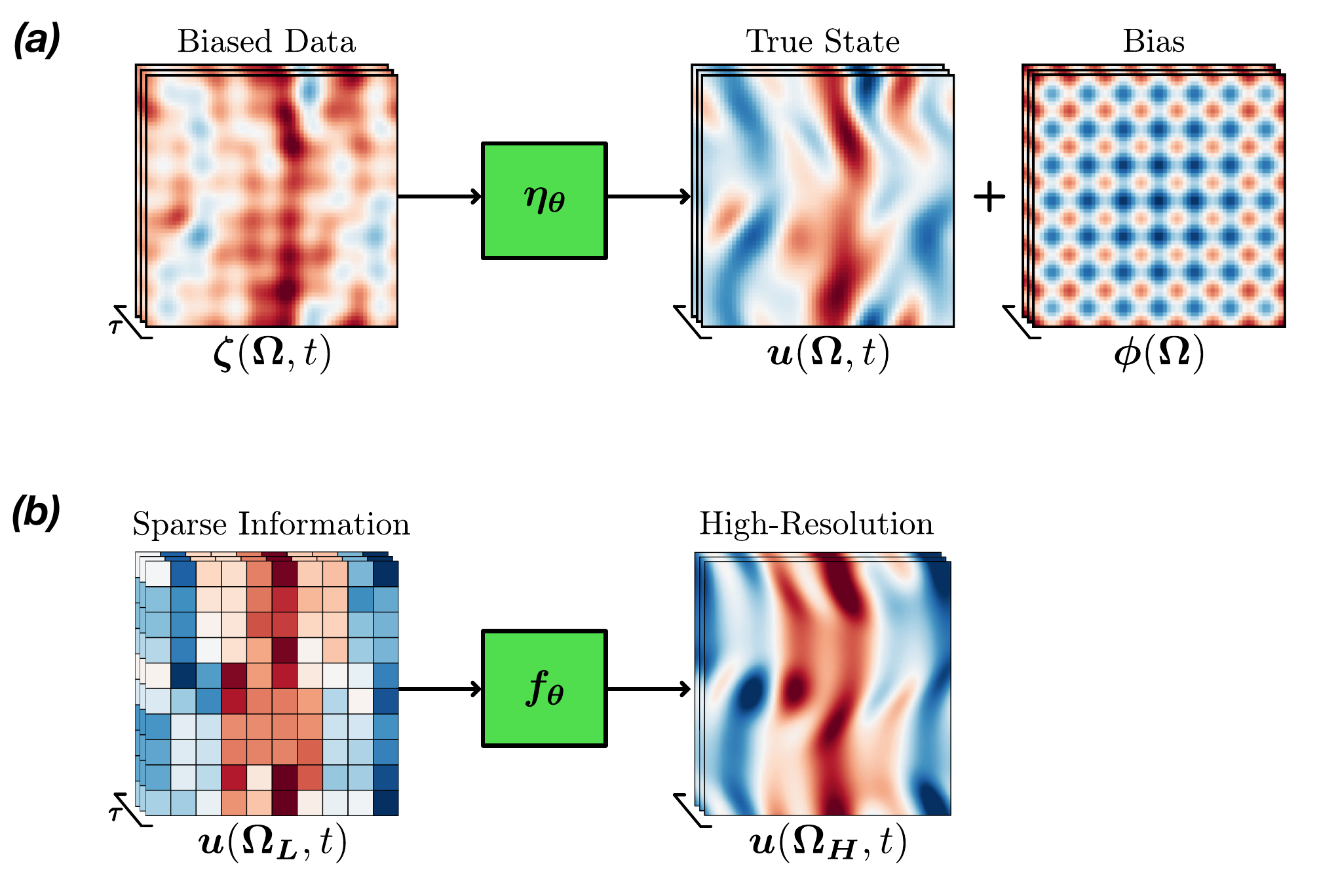}
    \caption{
        Inverse problems investigated in this paper. 
        (a) Uncovering solutions from biased data. The model ${\bm {\eta_\theta}}$ is responsible for recovering the  solution (true state), ${\bm u}({\bm \Omega}, t)$, from the biased data, ${\bm \zeta}(\bm{\Omega}, t)$. The bias (systematic error), ${\bm \phi}({\bm x})$, is the difference between the biased data and the solution. 
        (b) Reconstructing a solution from sparse information. The model ${\bm {f_\theta}}$ is responsible for mapping the sparse field ${\bm u}({\bm \Omega_L}, t)$ to the high-resolution field ${\bm u}({\bm \Omega_H}, t)$. The term $\tau$ in both cases denotes the number of contiguous time-steps passed to the network, required for computing temporal derivatives. An explanation of the proposed physics-constrained convolutional neural network (PC-CNN), which is the ansatz for both mappings ${\bm \eta_{\theta}}$ and ${\bm f_{\theta}}$, is provided in Sec.~\ref{sec:pc_cnn_overview}.
    }
    \label{fig:map_overview}
\end{figure*}

We first consider the inverse problem of uncovering the solution of a PDE~\eqref{eqn:dynamical_system} from biased data. We assume that we have biased information on the system's state
\begin{equation}
    \bm{\zeta}(\bm{x}, t) = {\bm u}(\bm{x}, t) + {\bm \phi}({\bm x}), \label{eqn:corruption_definition}
\end{equation}
where ${\bm \phi}$ is a bias, which in practice, is spatially varying (systematic error) that can be caused by miscalibrated sensors~\cite{sciacchitano2015collaborative}, or modelling assumptions~\cite{novoa2022real}. The quantity $\bm{\zeta}(\bm{x}, t)$ in Eq.~\eqref{eqn:corruption_definition} constitutes the biased dataset, which is not a solution of the PDE~\eqref{eqn:dynamical_system}, i.e.
\begin{align}
    \mathcal{R}({\bm \zeta}({\bm x}, t); {\bm p}) \neq 0. 
\end{align}
Given the biased information, ${\bm \zeta}$, our goal is to uncover the correct solution to the governing equations, ${\bm u}$, which is referred to as the true state. Computationally, we seek a mapping, ${\bm {\eta_\theta}}$, such that 
\begin{equation}\label{eqn:bias_map}
    {\bm {\eta_\theta}}: {\bm \zeta}({\bm \Omega}, t) \mapsto {\bm u}({\bm \Omega}, t),
\end{equation}
where the domain $\Omega$ is discretised on a uniform, structured grid ${\bm \Omega} \subset \mathbb{R}^{N^n}$. We choose the mapping ${\bm {\eta_\theta}}$ to be a convolutional neural network, which depends on trainable parameters ${\bm \theta}$ (Sec.~\ref{sec:cnn_overview}). 
The nature of the partial differential equations has a computational implication. For a linear partial differential equation, the residual is an explicit function of the bias, i.e. $\mathcal{R}({{\bm \zeta}({\bm x}, t)}; {\bm p}) = \mathcal{R}({{\bm \phi}({\bm x}, t)};{\bm p})$. This does not hold true for nonlinear systems, which makes the problem of removing the bias  more challenging. This is discussed in Sec.~\ref{sec:results:reconstruction}.
Figure~\ref{fig:map_overview}(a) provides an overview of the inverse problem of uncovering solutions from biased data. 

\subsection{Reconstruction from sparse information} \label{sec:reconstruction_methodology}

We formalise the inverse problem of reconstructing the solution from sparse information. Given sparse information, ${\bm u}({\bm \Omega_L}, t)$, we aim to reconstruct the solution to the partial differential equation on a fine grid, ${\bm u}({\bm \Omega_H}, t)$, where the domain $\Omega$ is discretised on low- and high-resolution uniform grids, ${\bm \Omega_L} \subset \mathbb{R}^{N^n}$ and ${\bm \Omega_H} \subset \mathbb{R}^{M^n}$, respectively, such that  ${\bm \Omega_L} \cap {\bm \Omega_H} = {\bm \Omega_L}$; $M = \kappa N$; and $\kappa \in \mathbb{N}^{+}$ is the scale factor. Our objective is to find ${\bm u}({\bm \Omega_H}, t)$ given the sparse information ${\bm u}({\bm \Omega_L}, t)$. We achieve this by seeking a parameterised mapping ${\bm {f_\theta}}$ such that
\begin{equation}\label{eqn:reconstruction_map}
    {\bm {f_\theta}}: {\bm u}({\bm \Omega_L}, t) \rightarrow {\bm u}({\bm \Omega_H}, t), 
\end{equation}
where the mapping ${\bm {f_\theta}}$ is a convolutional neural network parameterised by ${\bm \theta}$\footnote{With a slight abuse of notation, ${\bm \theta}$ denotes the trainable parameters for Eqs.~(\ref{eqn:bias_map},~\ref{eqn:reconstruction_map})}. We consider the solution ${\bm u}$ to be discretised with $N_t$ times steps. Figure~\ref{fig:map_overview}(b) provides an overview of the reconstruction task.

\section{Convolutional neural networks} \label{sec:cnn_overview}

The mappings ${\bm {\eta_\theta}}$ and ${\bm {f_\theta}}$ (Eqs.~(\ref{eqn:bias_map},~\ref{eqn:reconstruction_map}) respectively) are design choices. For data from spatiotemporal partial differential equations, we look for a tool that:
\textit{(i)} can, in principle, learn any continuous functions, i.e. is a universal approximator; 
\textit{(ii)} learn spatial correlations, as partial differential equations are defined by local operators;
\textit{(iii)} do not violate the conservation laws imposed by the PDE~\eqref{eqn:dynamical_system}; and
\textit{(iv)} can handle sequential data for the temporal dynamics.  
In this paper, we choose convolutional neural networks~\citep{lecun1998ConvolutionalNetworksImages}, which are suitable tools for learning datasets of partial differential equations, from fluid mechanics~\cite{murata2020NonlinearModeDecomposition, Gao2021, kelshaw2023SuperresolvingSparseObservations} to heat transfer~\cite{kim2020PredictionTurbulentHeat, edalatifar2021UsingDeepLearning}.
Convolutional neural networks naturally fulfil requirements \textit{(i, ii)}, but do not fulfil \textit{(iii, iv)}. In this paper, we propose the physics-constrained convolutional neural network (PC-CNN) to fulfil all requirements \textit{(i--iv)}. A convolutional neural network, ${\bm k}_{\bm \theta}$, is a composition of functions
\begin{equation}\label{eqn:cnn_composition}
    {\bm {k_\theta}} = 
        {\bm k}^{Q}_{\bm \theta_Q} \circ \cdots \circ h({\bm k}^{2}_{\bm \theta_2}) \circ h({\bm k}^{1}_{\bm \theta_1}), 
\end{equation}
where ${\bm k}^i_{{\bm \theta}_i}$ denote discrete convolutional layers; $h$ is an element-wise nonlinear activation function, which increases the expressive power of the network; and $Q$ is the number of layers. The convolutional layers\footnote{CNNs  work on structured grids. To deal with unstructured grids and complex domains, the data can be interpolated onto a structured grid and the boundary can be padded. The extension to unstructured grids and intricate domain is beyond the scope of this paper, which is left for future work.} are responsible for the discrete operation ${\bm k}_{\bm \theta}: ({\bm x}; \bm{w}, \bm{b}) \mapsto {\bm w} \ast {\bm d} + {\bm b}$, where ${\bm \theta} = ({\bm w}, {\bm b})$, $\ast$ is the convolution operation, ${\bm x}$ is the data, ${\bm w}$ are the trainable weights of the kernel, and ${\bm b}$ are the trainable biases. As the kernel operates locally around each pixel, information is leveraged from the surrounding grid cells. For a pedagogical explanation about CNNs the reader is referred to~\citep{magri2023notes}.
%

\section{Physics-constrained convolutional neural networks (PC-CNN)} \label{sec:pc_cnn_overview}

We propose a physics-constrained convolutional neural network (PC-CNN) with time-windowing to allow for processing of sequential data. In this section, we describe the loss functions for the tasks of uncovering solutions from biased data and reconstruction from sparse information (Secs.~\ref{sec:bias_removal_methodology},~\ref{sec:reconstruction_methodology} respectively), and the time windowing for computing the time derivatives required in the physics-constraint. 

\subsection{Losses for training} \label{sec:loss_terms}
The mappings ${\bm {\eta_\theta}}$ and ${\bm {f_\theta}}$ are parameterised as convolutional neural networks with the ansatz of Eq.~\eqref{eqn:cnn_composition}. Training these networks is an optimisation problem, which seeks an optimal set of parameters ${\bm \theta}^\ast$ such that 
\begin{align} \label{eqn:optimisation_problem}
\begin{split}
    &{\bm \theta}^\ast = \argmin_{\bm \theta} \mathcal{L}_{\bm \theta} 
    \\ &\text{where} \quad
    \mathcal{L}_{\bm \theta} = \lambda_D\mathcal{L}_{\mathcal{D}} +  \lambda_P \mathcal{L}_{\mathcal{P}} + \lambda_{C} \mathcal{L}_{C},
\end{split}
\end{align} 
where the loss function $\mathcal{L}_{\bm \theta}$ consists of: 
$\mathcal{L}_{D}$, which quantifies the error between the data and the prediction ("data loss"); 
$\mathcal{L}_{P}$, which is responsible for quantifying how unphysical the predictions are ("physics loss"); and 
$\mathcal{L}_{C}$, which embeds any further required constraints ("constraint loss").
Each loss is weighted by a non-negative regularization factor, $\lambda_{(\cdot)}$, which is an empirical hyperparameter. The definitions of the losses and regularization factors are task dependent; that is, we must design a suitable loss function for each inverse problem we solve.

\subsubsection{Losses for uncovering solutions from biased data} \label{sec:loss:removing_bias}

Given biased data ${\bm \zeta}({\bm \Omega}, t)$, we define the loss terms for the task of uncovering solutions as 
\begin{align} \label{eqn:bias_removal_loss_terms}
   \mathcal{L}_{D} &= \frac{1}{N_{t}} \sum_{i = 0}^{N_t - 1} \big\lVert
        {\bm {\eta_\theta}}({\bm \zeta}(\partial {\bm \Omega}, t_i)) - {\bm u}(\partial {\bm \Omega}, t_i)
    \big\rVert_{\partial {\bm \Omega}}^{2}, \\
        \mathcal{L}_{P} &= \frac{1}{N_{t}} \sum_{i = 0}^{N_t - 1} \big\lVert
        \mathcal{R}\big({\bm {\eta_\theta}}({\bm \zeta}({\bm \Omega}, t_i)); {\bm p} \big)
   \big\rVert_{{\bm \Omega}}^{2}, \\
   \mathcal{L}_{C} &= \frac{1}{N_{t}} \sum_{i = 0}^{N_t - 1} \big\lVert
        \partial_t \big[ {\bm \zeta}({\bm \Omega}, t_i) - {\bm {\eta_\theta}}({\bm \zeta}({\bm \Omega}, t_i)) \big]
   \big\rVert_{{\bm \Omega}}^{2},
\end{align}
where $\partial {\bm \Omega}$ denotes boundary points of the grid, and $\lVert \blank \rVert_{{\bm \Omega}}$ is the $\ell^2$-norm over the discretised domain.  
We impose the prior knowledge that we have on the dynamical system by defining the physics loss, $\mathcal{L}_{P}$, which penalises network parameters that yield predictions that violate the governing equations~\eqref{eqn:dynamical_system}\footnote{For example, in turbulence, the fluid mass and momentum must be in balance with mass sources, and forces, respectively. Thus, Eq.~\eqref{eqn:dynamical_system} represents conservation laws.}. Mathematically, this means that in absence of observations (i.e., data), the physics loss $\mathcal{L}_{P}$ alone does not provide a unique solution. By augmenting the physics loss with the data loss, $\mathcal{L}_{D}$ (see Eq.~\eqref{eqn:optimisation_problem}), we ensure that network realisations conform to the observations (data) on the boundaries whilst fulfilling the governing equations, e.g. conservation laws.
In the case of this inverse problem, the constraint loss $\mathcal{L}_{\bm C}$ prevents the bias ${\bm \phi} = {\bm \phi}(x)$ from temporally changing. Penalising network realisations that do not provide a unique bias helps stabilise training and drive predictions away from trivial solutions, such as the stationary solution ${\bm u}({\bm \Omega}, t) = 0$. Although we could use collocation points within the domain (as well as on the boundary), we show that our method can uncover the solution of the PDE with samples on the boundary only. Spatial sparsity could be increased by only using a subset of samples on the boundary, but for this task we use information from all the boundary nodes.

\subsubsection{Losses for reconstruction from sparse information} \label{sec:loss:reconstruction}

Given low-resolution observations ${\bm u}({\bm \Omega_L}, t)$, we define the loss terms for the reconstruction from sparse information as 
\begin{equation} \label{eqn:reconstruction_loss_terms}
\begin{aligned}[c]
    \mathcal{L}_{D} &= \frac{1}{N_t} \sum_{i = 0}^{N_t - 1} \lVert 
        \corestr{{\bm {f_\theta}}({\bm u}({\bm \Omega_L}, t_i))}{{\bm \Omega_L}} - {\bm u}({\bm \Omega_L}, t_i)
    \rVert_{{\bm \Omega_L}}^{2}, \\
    \mathcal{L}_{P} &= \frac{1}{N_t} \sum_{i = 0}^{N_t - 1} \lVert 
       \mathcal{R}({\bm {f_\theta}}({\bm u}({\bm \Omega_L}, t_i)); {\bm p})
    \rVert_{{\bm \Omega_H}}^{2},
\end{aligned}
\end{equation}
where $\corestr{{\bm {f_\theta}}(\blank)}{{\bm \Omega_L}}$ denotes the corestriction of ${\bm \Omega_H}$ on ${\bm \Omega_L}$. In order to find an optimal set of parameters ${\bm \theta}^\ast$, the loss is designed to regularise predictions that do not conform to the desired output. Given sparse information ${\bm u}({\bm \Omega_L}, t)$, the data loss, $\mathcal{L}_{D}$, is defined to minimise the distance between known sparse data,  ${\bm u}({\bm \Omega_L}, t)$, and their corresponding predictions on the high-resolution grid, $\corestr{{\bm {f_\theta}}({\bm u}({\bm \Omega_L}, t))}{{\bm \Omega_L}}$. 
Crucially, as consequence of the proposed training objective, we do not need the high-resolution field as labelled dataset, which is required with conventional super-resolution methods~\citep{Freeman2002, Dong2014, vincent2010StackedDenoisingAutoencoders}.

\subsection{Time-windowing of the physics loss}

\begin{figure*}[ht]
    \centering
    \includegraphics[width=\linewidth]{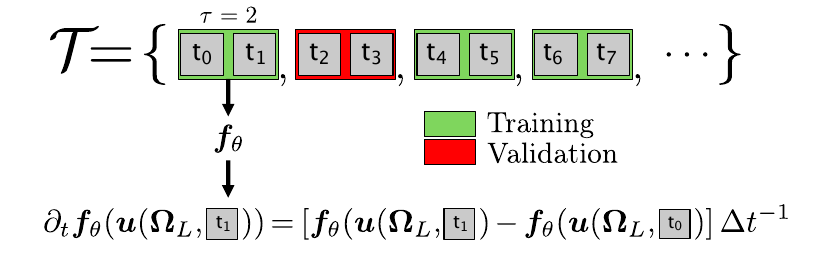}
    \caption{
        Time-windowing scheme. Time-steps are first grouped into non-overlapping subsets of successive elements of length $\tau$. Each of these subsets can be taken for either training, or validation. Subsets are treated as minibatches and passed through the network to evaluate their output. The temporal derivative is then approximated using a forward-Euler approximation across adjacent time-steps.
    }
    \label{fig:time-windowing-scheme}
\end{figure*}

Computing the residual of a partial differential equation is a temporal task, as shown in Eq.~\eqref{eqn:dynamical_system}. We propose a simple time-windowing approach to allow the network to account for the temporal nature of the data. This windowing approach provides a means to compute the time-derivative $\partial_t {\bm u}$ required for the physics loss $\mathcal{L}_{P}$. The network takes time-windowed samples in as inputs, each sample consisting of $\tau$ sequential time-steps. Precisely, we first group all time-steps into non-overlapping sets of $\tau$ sequential time-steps to form the set $\mathcal{T} = \{ [t_{i\tau}, \dots, t_{(i + 1)\tau - 1}] \}_{i=0}^{\nicefrac{N_t}{\tau}}$. Elements of this set are treated as minibatches and are passed through the network to evaluate their output on which the physics loss is computed. The time-derivative $\partial_t {\bm u}$ is computed by applying a forward-Euler approximation across successive time-steps. Figure~\ref{fig:time-windowing-scheme} provides an overview of the time-batching scheme and how elements of each batch are used to compute the temporal derivative.

Employing this time-windowing scheme, for the task of reconstructing from sparse information, the physics loss is 
\begin{equation}
    \mathcal{L}_{P} = \frac{1}{(\tau - 1) \lvert \mathcal{T} \rvert} \sum_{t \in \mathcal{T}} \sum_{i = 0}^{\tau - 1}
    \big\lVert 
            \partial_{t} {\bm {f_\theta}}({\bm u}({\bm \Omega_L}, t_{i + 1}))
            - 
            \mathcal{N}({\bm {f_\theta}}({\bm u}({\bm \Omega_L}, t_{i + 1})); {\bm p})
    \big\rVert_{{\bm \Omega_H}}^{2}, 
\end{equation}
whereas for the task of uncovering solutions from biased data, the physics loss is
\begin{equation}
    \begin{split}
    \mathcal{L}_{P} = \frac{1}{(\tau - 1) N_{t}} \sum_{t \in \mathcal{T}} \sum_{i = 0}^{\tau - 1}
    \big\lVert 
        \partial_{t} {\bm {\eta_\theta}}({\bm \Omega}, t_{i + 1}))
    - 
        \mathcal{N}({\bm {\eta_\theta}}({\bm \zeta}({\bm \Omega}, t_{i + 1})); {\bm p})
    \big\rVert_{{\bm \Omega}}^{2},
        \end{split}
\end{equation}
where $\lvert \mathcal{T} \rvert$ denotes the cardinality of the set, or the number of minibatches\footnote{Note, the updated versions of the physics losses fundamentally represent the same objective, but differ slightly in their definition.}.
Predicting the high-resolution field for $\tau$ sequential time-steps allows us to obtain the residual for the predictions in a temporally local sense, i.e., we compute the derivatives across discrete time windows rather than the entire simulation domain. The networks ${\bm {f_\theta}}$ and ${\bm {\eta_\theta}}$  are augmented to operate on these time windows, which vectorises the operations over the time-window.
For a fixed quantity of training data, the choice of $\tau$ introduces a trade-off between the number of input samples $N_t$, and the size of each time-window $\tau$. A larger value of $\tau$ corresponds to fewer time windows. Limiting the number of time windows used for training  has an adverse effect on the ability of the model to generalise; the information content of contiguous timesteps is scarcer than taking timesteps uniformly across the time domain. Although evaluating the residual across large time windows promotes numerical stability, this  smooths the gradients computed in backpropagation, and makes the model more difficult to train, especially in the case of chaotic systems where the dynamics might vary erratically. Upon conducting a hyperparameter study, we take a value $\tau = 2$, which is also the minimum window size for computing the temporal derivative of the physics loss by finite difference. We find that this is sufficient for training the network whilst simultaneously maximising the number of independent samples used for training. 
To avoid duplication of the data in the training set, we ensure that all samples are at least $\tau$ time-steps apart so that independent time-windows are guaranteed to not contain any overlaps.
The remaining loss terms operate in the same manner, i.e. over the time window as well as the conventional batch dimension.

\section{Synthetic data generation} \label{sec:spectral_method}
We utilise a differentiable pseudospectral spatial discretisation to solve the partial differential equations of this paper~\cite{kolsol2022}. The solution is computed on the spectral grid ${\bm {\hat \Omega}}_k \in \mathbb{Z}^{K}$, where $K$ is the number of wavenumbers. This spectral discretisation enforces periodic boundary conditions on the domain boundary, $\partial \Omega$. A solution is produced by time-integration of the dynamical system with the explicit forward-Euler scheme, in which we choose the timestep $\Delta t$ to satisfy the Courant-Friedrichs-Lewy (CFL) condition~\cite{canuto1988SpectralMethodsFluid}. The initial conditions in each case are generated using the equation
\begin{equation} \label{eqn:initial_condition}
\begin{split}
    {\bm u}({\bm {\hat \Omega}}_k, 0) = \frac{\iota e^{2\pi i {\bm \epsilon}}}{\sigma \sqrt{2\pi}} e^{-\frac{1}{2} \left( \frac{\lvert {\bm {\hat \Omega}}_k \rvert}{\sigma} \right)^2}
        \\ \qquad \text{with} \quad {\bm \epsilon}_i \sim N(0, 1),
        \end{split}
\end{equation}
where $\iota$ denotes the magnitude; $\sigma$ denotes the standard deviation; and ${\bm \epsilon}_i$ is a sample from a unit normal distribution. Equation~\eqref{eqn:initial_condition} produces a pseudo-random field scaled by the magnitude of the wavenumber $\lvert {\bm {\hat \Omega}}_k \rvert$, where magnitude is computed pointwise for each wavenumber in the spectral grid. This scaling ensures that the resultant field has spatial structures of varying lengthscale. We take $\iota = 10, \sigma = 1.2$ for all simulations in order to provide an initial condition, which provides numerical stability.

As a consequence of the Nyquist-Shannon sampling criterion~\cite{canuto1988SpectralMethodsFluid}, the resolution of the spectral grid ${\bm {\hat \Omega}}_k$ places a lower bound on the spatial resolution. For a signal containing a maximal frequency $\omega_\text{max}$, the sampling frequency $\omega_s$ must satisfy the condition $\omega_\text{max} < \nicefrac{\omega_s}{2}$, therefore, we ensure that the spectral resolution satisfies $K \leq \nicefrac{N}{2}$. Violation of this condition induces spectral aliasing, in which energy content from frequencies exceeding the  Nyquist limit $\nicefrac{\omega_s}{2}$ is fed back to the low frequencies, which amplifies energy content unphysically~\citep{canuto1988SpectralMethodsFluid}. To prevent aliasing, in the task of uncovering solutions from biased data, we employ a spectral grid ${\bm {\hat \Omega}}_k \in \mathbb{Z}^{32 \times 32}$, which corresponds to the physical grid ${\bm \Omega} \in \mathbb{R}^{64 \times 64}$. For the task of reconstructing the solution from sparse data, we employ a high-resolution spectral grid ${\bm {\hat \Omega}}_k \in \mathbb{Z}^{35 \times 35}$, which corresponds to the physical grid ${\bm \Omega} \in \mathbb{R}^{70 \times 70}$. Approaching the Nyquist limit allows us to resolve the smallest structures possible without introducing spectral aliasing.

\subsection{Physics loss in the Fourier domain} \label{sec:fourier_loss}

The pseudospectral discretisation provides an efficient means to compute the differential operator $\mathcal{N}$, which allows us to evaluate the physics loss $\mathcal{L}_{P}$ in the Fourier domain. Computationally, we now evaluate the physics loss for the task of uncovering solutions from biased data as
\begin{equation}
\begin{split}
    \mathcal{L}_{P} = \frac{1}{(\tau - 1) \lvert \mathcal{T} \rvert} \sum_{t \in \mathcal{T}} \sum_{i=0}^{\tau - 1} \lVert
        \partial_t \hat{{\bm {\eta}}}_\theta({\bm \zeta}({\bm \Omega}, t_{i + 1}))
        - \hat{\mathcal{N}}(\hat{{\bm {\eta}}}_\theta({\bm \zeta}({\bm \Omega}, t_{i + 1})))
    \rVert_{\hat{\bm \Omega}_{k}}^{2},
    \end{split}
\end{equation}
and for the task of reconstruction from sparse data as
\begin{equation}
\begin{split}
    \mathcal{L}_{P} = \frac{1}{(\tau - 1) \lvert \mathcal{T} \rvert} \sum_{t \in \mathcal{T}} \sum_{i = 0}^{\tau - 1} \lVert
        \partial_t \hat{\bm {f_\theta}}({\bm u}({\bm \Omega_L}, t_{i + 1})) - \hat{\mathcal{N}}(\hat{\bm {f_\theta}}({\bm u}({\bm \Omega_L}, t_{i + 1})))
    \rVert_{\bm{\hat{\Omega}_{k}}}^{2},
\end{split}
\end{equation}
where $\hat{\cdot} = \mathcal{F}\{{\cdot}\}$ is the Fourier-transformed variable, and $\hat{\mathcal{N}}$ denotes the Fourier-transformed differential operator. The pseudospectral discretisation is fully differentiable, which allows us to numerically compute gradients with respect to the parameters of the network ${\bm \theta}$. Computing the loss $\mathcal{L}_{P}$ in the Fourier domain provides two advantages: \textit{(i)} periodic boundary conditions are naturally enforced, which enforces the  prior knowledge in the loss calculations; and \textit{(ii)} gradient calculations yield spectral accuracy. In contrast, a conventional finite differences approach requires a computational stencil, the spatial extent of which places an error bound on the gradient computation. This error bound is a function of the spatial resolution of the field. 
\section{Results on uncovering solutions from biased data} \label{sec:results:bias_removal}

We first introduce the mathematical parameterisation of the bias (i.e., systematic error).
Second, we discuss numerical details and performance metrics used for the analysis.
Third, we showcase results for three partial differential equations of increasing complexity: the linear convection-diffusion, nonlinear convection-diffusion (Burgers' equation~\cite{bateman1915, burgers1948}), and two-dimensional turbulent Navier-Stokes equations (the Kolmogorov flow~\cite{Fylladitakis2018}). The Navier-Stokes, which is a spatiotemporal chaotic system, will be used for the task of reconstruction from sparse information in Sec.~\ref{sec:results:reconstruction}. 
Finally, we analyse the physical consistency of physics-constrained convolutional neural network (PC-CNN) predictions for the two-dimensional turbulent flow. 

\subsection{Parameterisation of the bias} \label{sec:bias_parameterisation}
Errors on predictions can broadly fall into two catergories: \textit{(i)} aleatoric errors, which are typically referred to as noise due to environmental and sensors' stochastic behaviour; and \textit{(ii)} epistemic uncertainties, which are typically referred to as bias. We parameterise the bias with a non-convex spatially varying function to model an epistemic error. We employ a modified Rastrigin parameterisation~\cite{Rastrigin1974}, which is commonly used in non-convex optimization benchmarks

\begin{equation} 
\label{eqn:parameterised_bias}
    {\bm \phi}({\bm x}; \mathcal{M}, k_\phi, u_{\text{max}}) =
    \frac{\mathcal{M} u_{\text{max}}}{2\pi^2 + 40} \left( 20 + \sum_{i=1}^{2} \left[ ({\bm x}_{i} - \pi)^{2} - 10 \cos(k_\phi ({\bm x}_{i} - \pi)) \right] \right)
\end{equation}
where $u_{\text{max}}$ is the maximum absolute velocity in the flow;  $\mathcal{M}$ is the relative magnitude of the bias;  $k_\phi$ is the wavenumber of the bias, and ${\bm x}_{1}$ and ${\bm x}_{2}$ are the streamwise and transversal coordinates respectively. Parameterising the bias with $\mathcal{M}$ and $k_\phi$ allows us to evaluate the performance of the methodology with respect to different degrees of non-convexity of the bias. Aleatoric noise removal is a fairly established field, and is not the focus of this task.

\subsection{Numerical details, data, and performance metrics} \label{sec:numerical_details_bias_removal}

We uncover the solutions of partial differential equations from biased data for three physical systems of increasing complexity. Each system is solved via the pseudospectral discretisation as described in Sec.~\ref{sec:spectral_method}, producing a solution by time-integration using the Euler-forward method with timestep of $\Delta t = 5 \times 10^{-3}$, which is chosen to satisfy of the Courant-Friedrichs-Lewy (CFL) condition to promote numerical stability. 
Model training is performed with $1024$ training samples and $256$ validation samples, which are pseudo-randomly selected from the time-domain of the solution. Each sample contains $\tau = 2$ sequential time-steps. Following a hyperparameter study, the \texttt{adam} optimiser is employed~\cite{kingma2015} with a learning rate of $3 \times 10^{-4}$. The weighting factors for the loss, as described in Eq.~\ref{eqn:optimisation_problem}, are taken as $\lambda_P = \lambda_C = 10^3$, which are empirically determined (through a hyperparameter search) to provide stable training. Details on the network architecture can be found in~\ref{app:network:picr}. Each model is trained for a total of $10^4$ epochs, which is chosen to  provide sufficient convergence. The accuracy of prediction is quantified by the relative error on the validation dataset 
\begin{equation} \label{eqn:error:bias_removal}
    e = 
    \sqrt{
        \frac{
            \sum_{t \in \mathcal{T}} \lVert {\bm u}(\bm{\Omega}, t) - {\bm {\eta_\theta}}({\bm \zeta}(\bm{\Omega}, t)) \rVert_{\bm{\Omega}}^{2}
        }{
            \sum_{t \in \mathcal{T}} \lVert {\bm u}(\bm{\Omega}, t) \rVert_{\bm{\Omega}}^{2}
        }
    }.
\end{equation}
This metric takes the magnitude of the solution into account, which allows the results from different systems to be compared. All experiments in this paper are run on a single Quadro RTX 8000 GPU. Numerical tests are run with different random initialisations (seeds) to assess the robustness of the network.

\subsection{The linear convection-diffusion equation} \label{sec:qualitative:linear}
The linear convection-diffusion equation is used to describe transport phenomena~\cite{majda1999simplified}
\begin{equation} \label{eqn:linear_cd}
    \partial_t {\bm u} + {\bm c}\cdot \nabla {\bm u} = {\nu} \Delta {\bm u}, 
\end{equation}
where $\nabla $ is the nabla operator; $\Delta$ is the Laplacian operator; ${\bm c}\equiv (c,c)$, where $c$ is the convective coefficient; and ${\nu}$ is the diffusion coefficient. The flow energy is subject to rapid decay because the solution quickly converges towards a fixed-point solution at ${\bm u}(\bm{\Omega}, t) = 0$, notably in the case where $\nicefrac{\nu}{\bm{c}}$ is large. In the presence of the fixed-point solution, we observe ${\bm \zeta}({\bm \Omega}, t) = {\bm \phi}({\bm \Omega})$, which is a trivial case for identification and removal of the bias. In order to avoid rapid convergence to the fixed-point solution, we take ${c} = 1.0, {\nu} = \nicefrac{1}{500}$ as the coefficients, producing a convective-dominant solution. A snapshot of the results for $k_\phi = 3, \mathcal{M} = 0.5$ is shown in Figure~\ref{fig:snapshot_results}(a). There is a marked difference between the biased data ${\bm \zeta}({\bm \Omega}, t)$ in panel \textit{(i)} and the true state ${\bm u}({\bm \Omega}, t)$ in panel \textit{(ii)}, most notably in the magnitude of the field. Network predictions ${\bm {\eta_\theta}}({\bm \zeta}({\bm \Omega}, t))$ in panel \textit{(iii)} uncover the solution to the partial differential equation with a relative error of $e = 6.612 \times 10^{-2}$ on the validation set. 

\subsection{Nonlinear convection-diffusion equation} \label{sec:qualitative:nonlinear}
The nonlinear convection-diffusion equation, also known as Burgers' equation~\cite{burgers1948}, is
\begin{equation} \label{eqn:nonlinear_cd}
    \partial_t {\bm u} + \left( {\bm u} \cdot {\nabla} \right) {\bm u} = {\nu} \Delta {\bm u},
\end{equation}
where the nonlinearity lies in the convective term $\left( {\bm u} \cdot {\nabla} \right) {\bm u}$. The nonlinear convective term provides a further challenge: below a certain threshold, the  velocity interactions lead to further energy decay. The kinematic viscosity is set to ${\nu} = \nicefrac{1}{500}$, which produces a convective-dominant solution by time integration. In contrast to the linear convection-diffusion system (Sec.~\ref{sec:qualitative:linear}), the relationship between the dynamics of the biased state and the true state is more challenging. The introduction of nonlinearities in the differential operator revoke the linear relationship between the bias and data, i.e. $\mathcal{R}({{\bm \zeta}({\bm x}, t)}; {\bm p}) \neq \mathcal{R}({{\bm \phi}({\bm x}, t)}; {\bm p})$ as discussed in Sec.~\ref{sec:bias_removal_methodology}. Consequently, this increases the complexity of the physics loss term $\mathcal{L}_{P}$.

Figure~\ref{fig:snapshot_results}(b) shows a snapshot of results for $k_\phi = 5, \mathcal{M} = 0.5$. The true state ${\bm u}({\bm \Omega}, t)$ of the partial differential equation contains high-frequency spatial structures, shown in panel \textit{(ii)}, which are less prominent in the biased data ${\bm \zeta}({\bm \Omega}, t)$, shown in panel \textit{(i)}. Despite the introduction of a nonlinear differential operator, we demonstrate that the network retains the ability to recover the true state, as shown in panels \textit{(ii, iv)}, with a relative error on the validation set of $e = 6.791 \times 10^{-2}$.

\subsubsection{Two-dimensional turbulent flow} \label{sec:qualitative:kolmogorov}

Finally, we consider a spatiotemporally chaotic flow, which is governed by the incompressible Navier-Stokes equations evaluated on a two-dimensional domain with periodic boundary conditions, which is also known as the Kolmogorov flow~\cite{Fylladitakis2018}. The partial differential equations are expressions of the mass and momentum conservation, respectively 
\begin{align} \label{eqn:kolmogorov}
\begin{split}
    {\nabla} \cdot {\bm u} &= 0, \\
    \partial_t {\bm u} + \left( {\bm u} \cdot {\nabla} \right) {\bm u} &= - {\nabla} p + { \nu} \Delta {\bm u} + {\bm g}.
\end{split}
\end{align}
where $p$ is the pressure field; and ${\bm g}$ is a body forcing, which enables the dynamics to be sustained by ensuring that the flow energy does not dissipate. The flow density is constant and assumed to be unity, i.e. the flow is incompressible.
The use of a pseudospectral discretisation allows us to eliminate the pressure term and handle the continuity constraint implicitly, as shown in the spectral representation of the Navier-Stokes equations in~\ref{app:pseudospectral_discretisation}. The spatiotemporal dynamics are chaotic at the prescribed viscosity ${\nu} = \nicefrac{1}{42}$. The periodic forcing is ${\bm g}(\bm{x}) = (\sin{(4\bm{x_2}), 0})$. In order to remove the transient and focus on a statistically stationary regime, the transient time series up to $T_t = 180.0$ is discarded.

\begin{figure*}[!htb]
    \centering
         \includegraphics[width=0.8\linewidth]{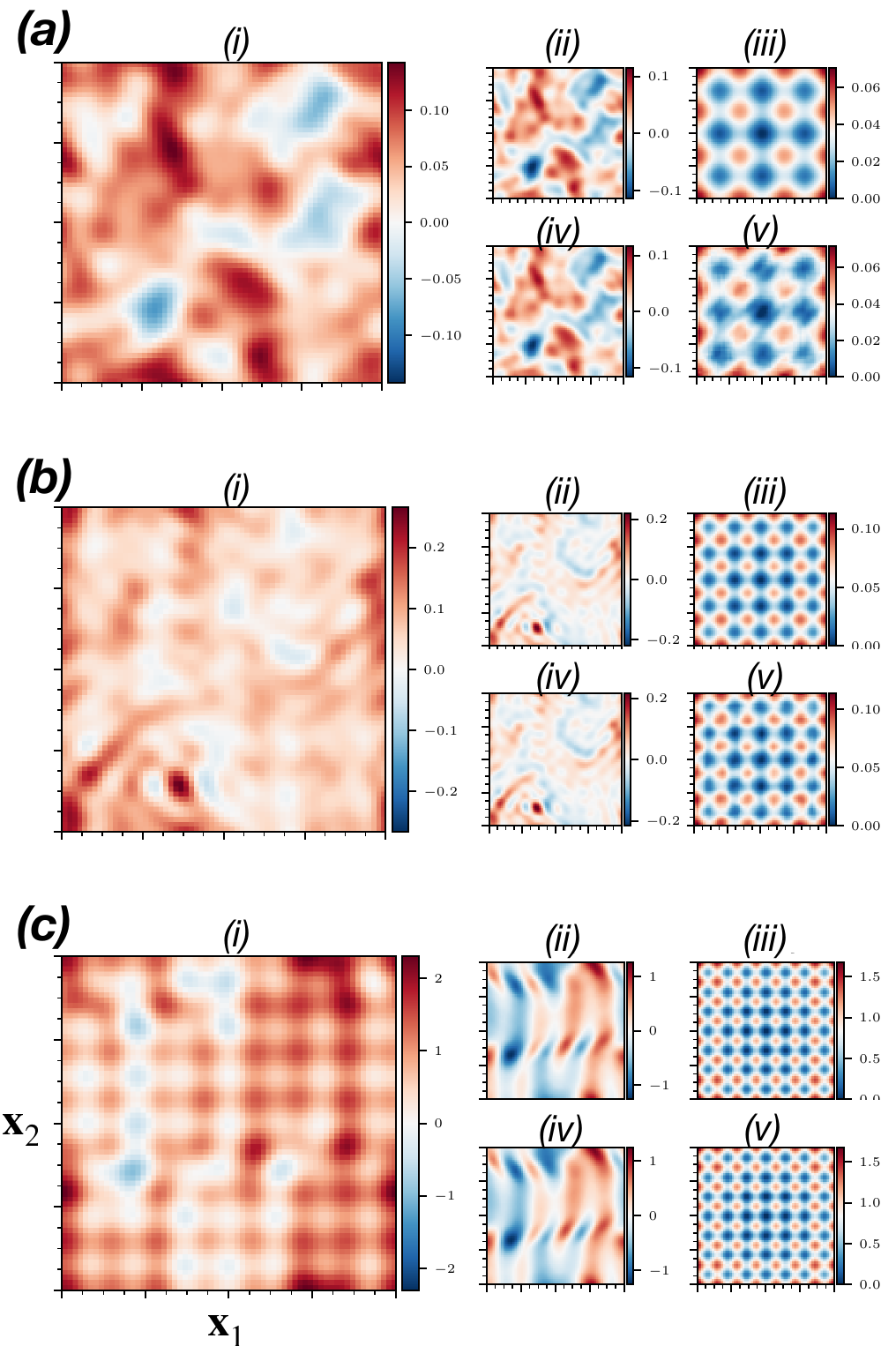}
        \caption{
            Uncovering solutions from biased data. (a) Linear convection-diffusion case $k_\phi = 3 $ and $\mathcal{M}=0.5$. (b) Nonlinear convection-diffusion case with $k_\phi = 5$ and $\mathcal{M}=0.5$. (c) Two-dimensional turbulent flow case with $[k_\phi = 7$ and $\mathcal{M}=0.5$. Panel \textit{(i)} shows the biased data, ${\bm \zeta}$; \textit{(ii)} shows the true state, ${\bm u}$, which we wish to uncover from the biased data; \textit{(iii)} shows the bias, which represents the bias (i.e., systematic error), ${\bm \phi}$;  \textit{(iv)} shows the network predictions, ${\bm \eta_\theta}$; and \textit{(v)} shows the predicted bias, ${\bm \zeta}$ - ${\bm \eta_\theta}$. 
            }
        \label{fig:snapshot_results}
\end{figure*}

A snapshot for the two-dimensional turbulent flow case is shown in Figure~\ref{fig:snapshot_results}(c) for $k_\phi = 7, \mathcal{M} = 0.5$. The bias field ${\bm \zeta}({\bm \Omega}, t)$ as shown in panel \textit{(i)} contains prominent, high-frequency spatial structures not present in the true state ${\bm u}({\bm \Omega}, t)$ shown in panel \textit{(ii)}. The biased field bares little resemblance to the true solution. In spite of the chaotic dynamics of the system, we demonstrate that network predictions ${\bm {\eta_\theta}}({\bm \zeta}({\bm \Omega}, t))$ shown in panel \textit{(iv)} successfully uncover the solution with a relative error on the  validation set of $e = 2.044 \times 10^{-2}$.

\FloatBarrier

\subsection{Robustness} \label{sec:bias_removal_robustness}
We analyse the robustness of the methodology by varying the wavenumber $k_\phi$ and magnitude $\mathcal{M}$ of the bias parameterization~\eqref{eqn:parameterised_bias}. Varying these parameters allows us to assess the degree to which the true state of a partial differential equation can be recovered when subjected to spatially varying bias with different degrees of non-convexity. To this end, we perform two parametric studies for each partial differential equation: 
\textit{(i)} $\mathcal{M} = 0.5$, $k_\phi \in \{ 1, 3, 5, 7, 9 \}$; and
\textit{(ii)} $k_\phi = 3$, $\mathcal{M} \in \{ 0.01, 0.1, 0.25, 0.5, 1.0 \}$.
For each case, we compute the relative error, $e$, between the predicted solution ${\bm {\eta_\theta}}({\bm \zeta}(\bm{\Omega}, t))$ and the true solution ${\bm u}(\bm{\Omega}, t)$, as defined in Eq.~\eqref{eqn:error:bias_removal}. We show the mean relative $\ell^2$-error for an ensemble of five computation for each test to ensure that the results are representative of the true performance. Empirically, we find that performance is robust to pseudo-random initialisation of network parameters with standard deviation of $\mathcal{O}(10^{-4})$ (result not shown).
We employ the same network setups described in Sec.~\ref{sec:numerical_details_bias_removal} with the same parameters for training. Assessing results using the same parameters for the three partial differential equations allows us to draw conclusion on the robustness of the methodology.

\begin{figure*}[!htb]
    \centering
    \includegraphics[width=\linewidth]{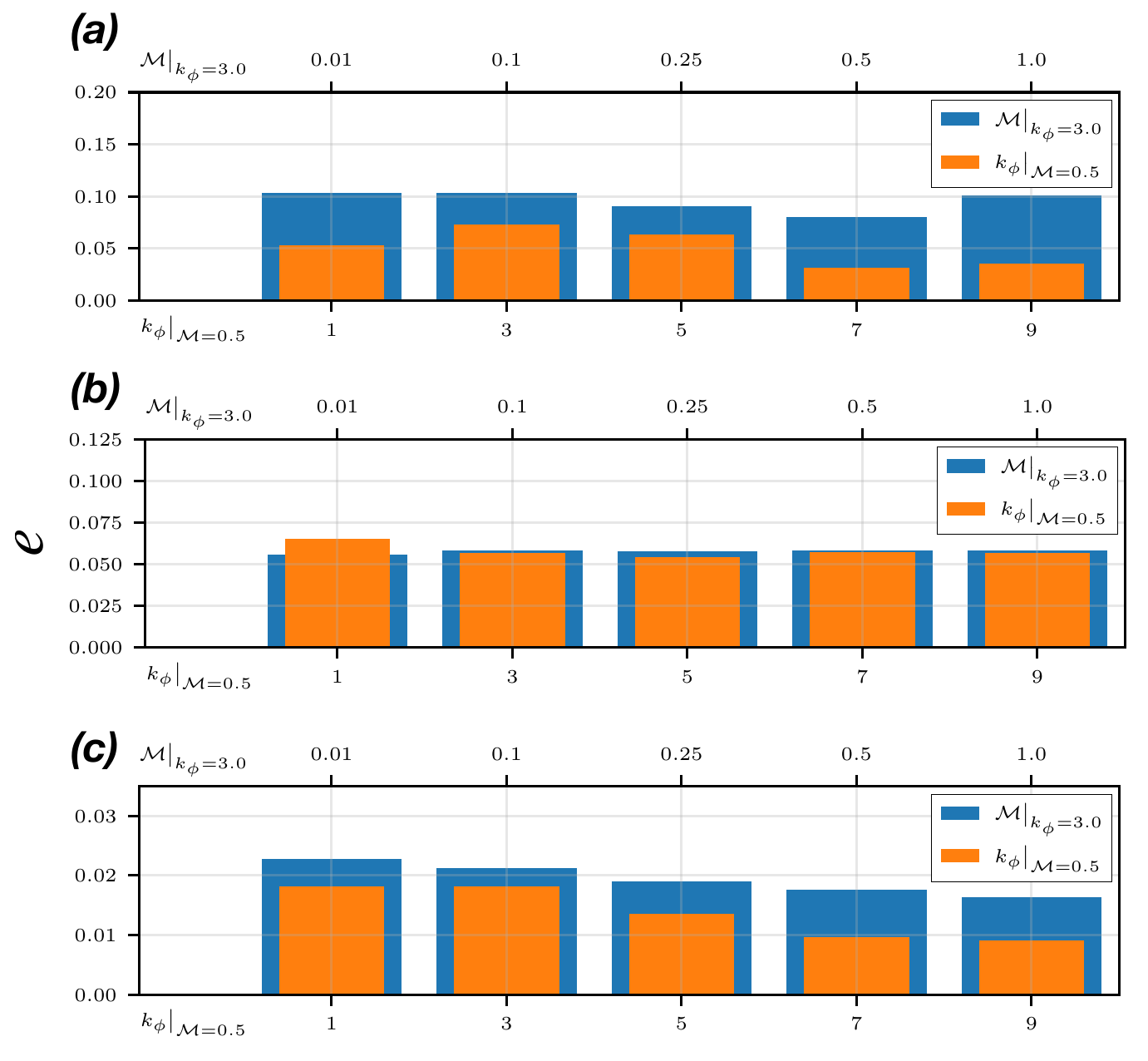}
    \caption{
       Unconvering solutions from biased data: robustness analysis through relative error, $e$. (a) Linear convection-diffusion case. (b) Nonlinear convection-diffusion case. (c) Two-dimensional turbulent flow case. Orange-bars denote results for case $\mathit{(i)}$: fixing the magnitude and varying the Rastrigin wavenumber. Blue-bars denote results for case $\mathit{(ii)}$: fixing the Rastrigin wavenumber and varying the magnitude.}
    \label{fig:parameterised_bias_results}
\end{figure*}

First, in the linear convection-diffusion problem as shown in Figure~\ref{fig:parameterised_bias_results}(a), we demonstrate that the relative error is largely independent of the form of parameterised bias. The relative magnitude $\mathcal{M}$ and Rastrigin wavenumber $k_\phi$ have small impact on the ability of the model to uncover the solution to the partial differential equation, with the performance being consistent across all cases. The model performs best for the case $k_\phi = 7, \mathcal{M} = 0.5$, achieving a relative error of $e = 2.568 \times 10^{-1}$.  Second, results for the nonlinear convection-diffusion case show marked improvement in comparison with the linear case, with errors for the numerical study shown in Figure~\ref{fig:parameterised_bias_results}(b). Despite the introduction of a nonlinear operator, the relative error is consistently lower regardless of the parameterisation of bias. The network remains equally capable of uncovering the true state across a wide range of modalities and magnitudes. Although the physics loss from the network predictions ${\bm {\eta_\theta}}({\bm \zeta}(\bm{\Omega}, t))$ no longer scales in a linear fashion, this is beneficial for the training dynamics. This is because the second order nonlinearity promotes convexity in the loss-surface, which is then exploited by our gradient-based optimisation approach. Third, the PC-CNN is able to uncover the true state of the turbulent flow. Results shown in Figure \ref{fig:parameterised_bias_results}(c) demonstrate an improvement in performance from the standard nonlinear convection-diffusion case. For both fixed Rastrigin wavenumber and magnitude, increasing the value of the parameter tends to decrease the relative error. The relative error in Figure \ref{fig:parameterised_bias_results}(c) decreases as the non-convexity of the bias increases. This is because it becomes increasingly simple to distinguish the bias from the underlying flow field if there is a larger discrepancy. We also emphasise that these tests were run with a fixed computational budget to ensure fair comparison -- with additional training, we observe a further decrease in relative error across all experiments. (We also ran tests with biases parameterized with multiple wavenumbers, an example of which is shown in Appendix~\ref{app:multi_freq}).

\FloatBarrier

\subsection{Physical consistency of the uncovered Navier-Stokes solutions} \label{sec:physical_consistency}

\begin{figure*}[!htb]
    \centering
    \includegraphics[width=\linewidth]{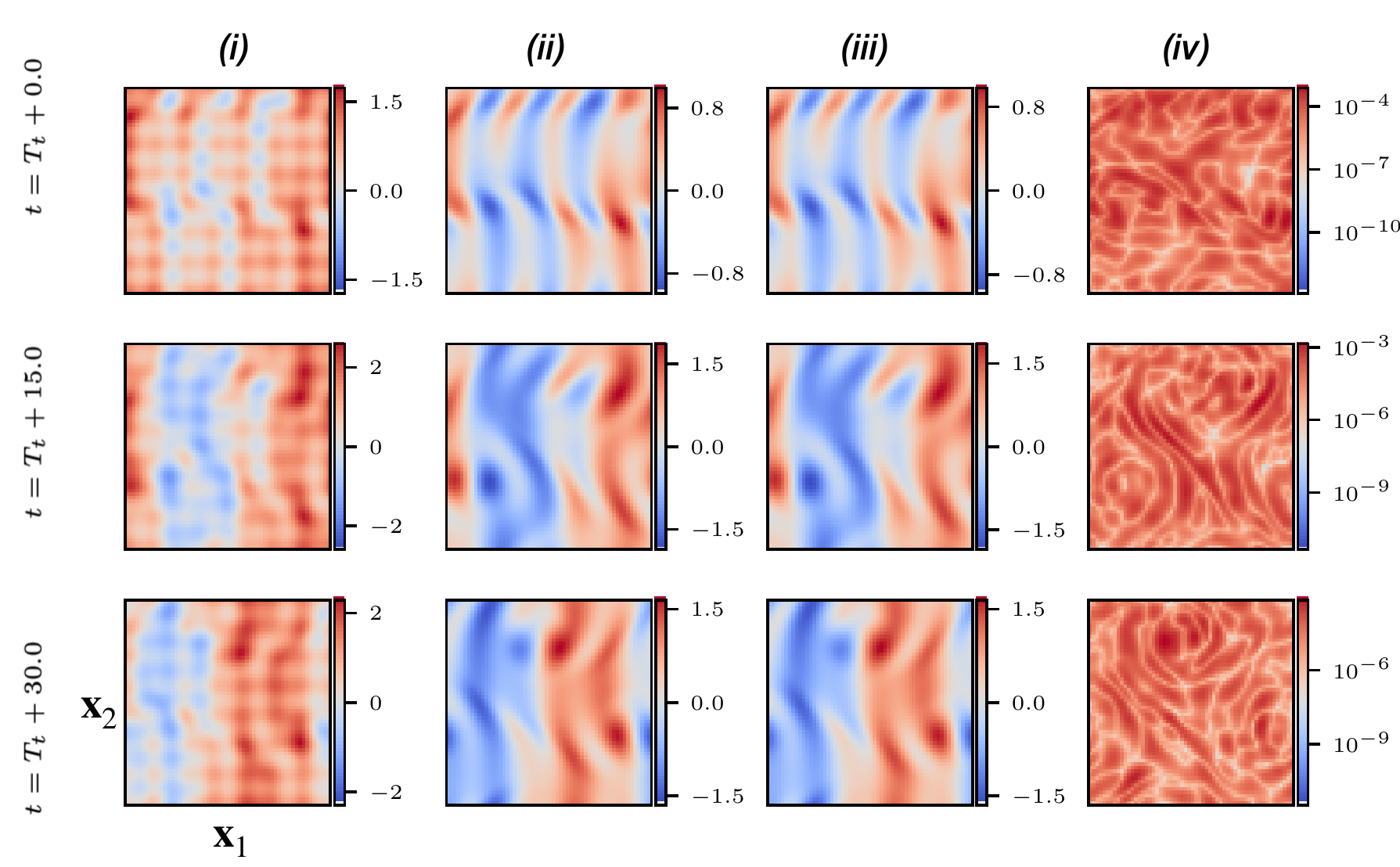}
    \caption{
        Uncovering solution from biased data. Temporal evolution of the two-dimensional turbulent flow with $[k_\phi = 7, \mathcal{M} = 0.5]$. $T_t$ denotes the length of the transient. \textit{(i)} Biased data, ${\bm \zeta}$; \textit{(ii)} true state, ${\bm u}$; \textit{(iii)} predicted solution, ${\bm \eta_\theta}$; and \textit{(iv)} squared error, $||{\bm \eta_\theta} - {\bm u}||^2$.}
    \label{fig:ts_kolmogorov}
\end{figure*}

Results in Sec.~\ref{sec:bias_removal_robustness} show  that, for a generic training setup, we are able to achieve small relative error for a variety of degrees of non-convexity. Because the two-dimensional turbulent flow case is chaotic, infinitesimal perturbations~$\sim O(\epsilon)$ exponentially grow in time, therefore, the residual $\mathcal{R}({\bm u}(\bm{\Omega}, t) + { \epsilon}(\bm{\Omega}, t); {\bm p}) \gg \mathcal{R}({\bm u}(\bm{\Omega}, t); {\bm p})$  where ${\epsilon}$ is a perturbation parameter. In this section, we analyse the physical properties of the solutions of the Navier-Stokes equation (Eq.~\ref{eqn:kolmogorov}) uncovered by the PC-CNN, ${\bm {\eta_\theta}}$. First, we show snapshots of the time evolution of the two-dimensional turbulent flow in Figure~\ref{fig:ts_kolmogorov}. These confirm that the model is learning a physical solution in time, as shown in the error on a log-scale in panels \textit{(iv)}. The parameterisation of the bias is fixed in this case with $k_\phi = 7, \mathcal{M} = 0.5$.

\begin{figure*}[!htb]
    \centering
    \includegraphics[width=\linewidth]{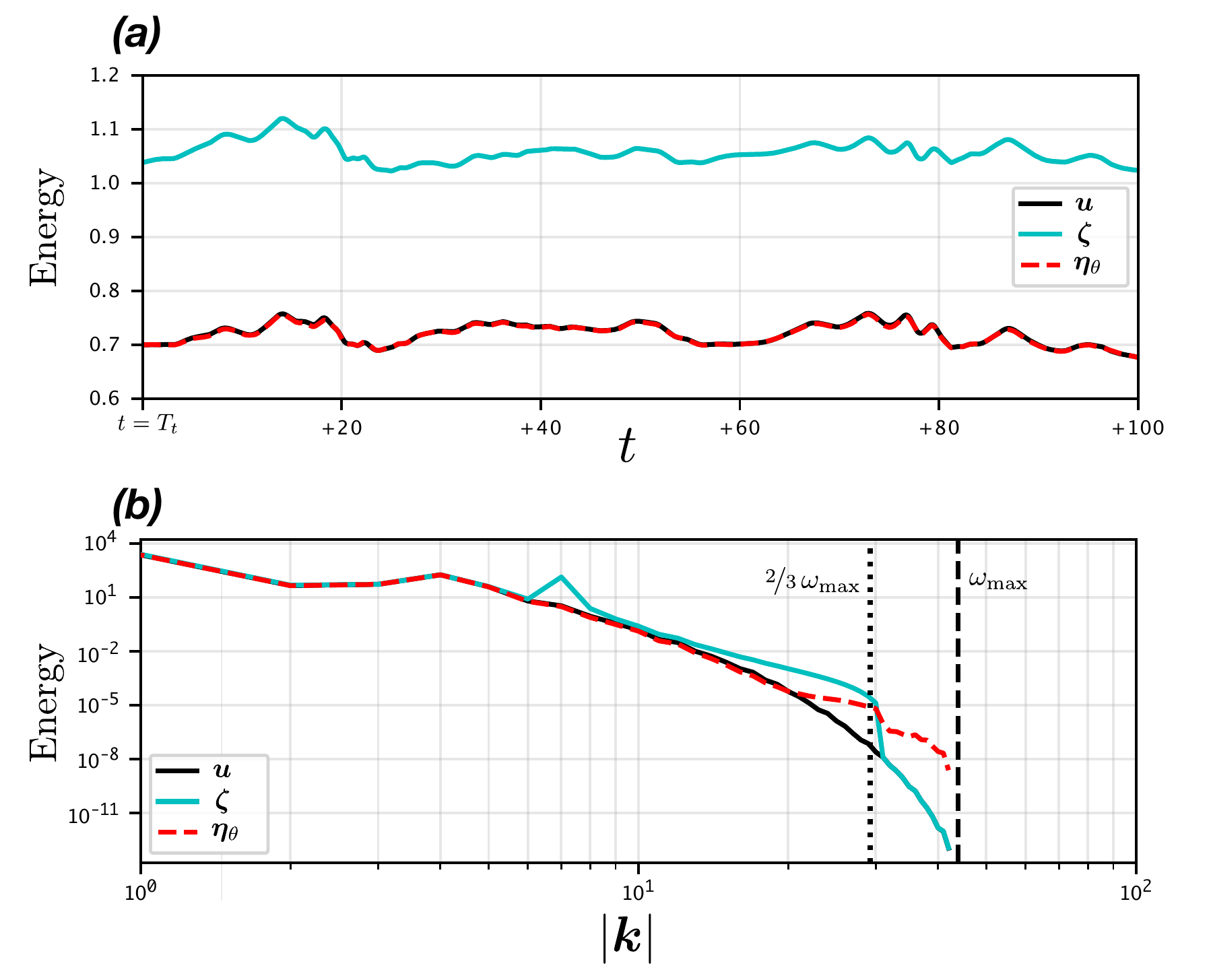}
    \caption{
        Uncovering solution from biased data: analysis of the solutions in the turbulent flow with $[k_\phi=7, \mathcal{M}=0.5]$. (a) Kinetic energy for the two-dimensional turbulent flow. (b) Energy spectrum for the two-dimensional turbulent flow. $T_t$ denotes the length of the transient.
        }
    \label{fig:physical_consistency_plot}
\end{figure*}

Second, we analyse the statistics of the solution. The mean kinetic energy of the flow at each timestep is directly affected by the introduction of the bias. In the case of our strictly positive bias ${\bm \phi}$ (Eq.~\eqref{eqn:corruption_definition}), the mean kinetic energy is increased at every point. Results in Figure~\ref{fig:physical_consistency_plot}(a) show the kinetic energy time series across the time-domain for: the true state ${\bm u}$; the biased data ${\bm \zeta}$; and the network's predictions ${\bm {\eta_\theta}}$. The chaotic nature of the solution can be observed from the erratic (but not random) evolution of the energy. We observe that the network predictions accurately reconstruct the kinetic energy trace across the simulation domain.

Third, we analyse the energy spectrum, which is characteristic of turbulent flows, in which the energy content decreases with the wavenumber. Introducing the bias at a particular wavenumber artificially alters the energy spectrum due to increased energy content. In Figure~\ref{fig:physical_consistency_plot}(b), we show the energy spectrum for the two-dimensional turbulent flow, where the unphysical increase in energy content is visible for $\lvert \bm{k} \rvert \geq 7$. Model predictions ${\bm {\eta_\theta}}({\bm \zeta}(\bm{\Omega}, t))$ correct the energy content for $\lvert \bm{k} \rvert < 21$, and successfully characterise and reproduce scales of turbulence. Although the energy content exponentially decreases with the spatial frequency, the true solution is correctly uncovered up to large wavenumbers, i.e. when aliasing occurs ($\lvert \bm{k} \rvert \gtrapprox 29$).

\FloatBarrier

\section{Results on reconstruction from sparse information} \label{sec:results:reconstruction} 

First, we discuss the generation of the sparse data. Next, we show the ability of the PC-CNN to infer the high-resolution solution of the partial differential equation on points that are not present in the training set.  
A high-resolution solution of the partial differential equation is generated on the grid ${\bm \Omega_H}$ prior to extracting a low-resolution grid ${\bm \Omega_L}$ with the scale factor of $\kappa = \nicefrac{N}{M}$ (Sec.~\ref{sec:reconstruction_methodology}). Both the solver and physics loss are discretised with $K = \nicefrac{N}{2}$ wavenumbers in the Fourier domain, which complies with the Nyquist-Shannon sampling criterion. The downsampling by $\kappa$ is performed by extracting the value of the solution at spatial locations ${\bm \Omega_L} \cap {\bm \Omega_H}$, i.e. a low-resolution representation of the high-resolution solution
\begin{equation}
    {\bm u}({\bm \Omega_L}, t) \triangleq \corestr{{\bm u}({\bm \Omega_H}, t)}{{\bm \Omega_L}}, 
\end{equation}
where $|^{{\bm \Omega_L}}$ is the corestriction which extracts points only in the relevant domain. While the data is sparse in space, a minimal temporal resolution is required for the computation of the time derivatives.

\subsection{Comparison with standard upsampling}
We show the results for a scale factor $\kappa = 7$, with results for further scale factors shown in~\ref{app:kappa_results}. Results are compared with interpolating upsampling methods, i.e. bi-linear, and bi-cubic interpolation to demonstrate the ability of the method. We quantify the accuracy by computing the relative error between the true solution, ${\bm u}({\bm \Omega_H}, t)$, and the corresponding network realisation, ${\bm {f_\theta}}({\bm u}({\bm \Omega_L}, t))$
\begin{equation} \label{eqn:error}
    e = 
    \sqrt{
        \frac{
            \sum_{t \in \mathcal{T}} \lVert {\bm u}(\bm{\Omega_H}, t) - {\bm {f_\theta}}({\bm u}(\bm{\Omega_L}, t)) \rVert_{\bm{\Omega_H}}^{2}
        }{
            \sum_{t \in \mathcal{T}} \lVert {\bm u}(\bm{\Omega_H}, t) \rVert_{\bm{\Omega_H}}^{2}
        }
    }.
\end{equation}
Upon discarding the transient, a solution ${\bm u}({\bm \Omega_H}, t) \in \mathbb{R}^{70 \times 70}$ is generated by time-integration over $12 \times 10^3$ time-steps, with $\Delta t = 1 \times 10^{-3}$.  We extract the low-resolution solution ${\bm u}({\bm \Omega_L}, t) \in \mathbb{R}^{10 \times 10}$. We extract $2048$ samples at random from the time-domain of the solution, each sample consisting of $\tau = 2$ consecutive time-steps. The \texttt{adam} optimiser~\citep{kingma2015} is employed for training with a learning rate of $3 \times 10^{-4}$. We take $\lambda_P = 10^{3}$ as the regularisation factor for the loss, as shown in Eq.~\eqref{eqn:optimisation_problem}, and train for a total of $10^{3}$ epochs, which is empirically determined to provide sufficient convergence, as per the results of a hyperparameter study (result not shown).

\begin{figure*}[!htb]
    \centering
    \includegraphics[width=\linewidth]{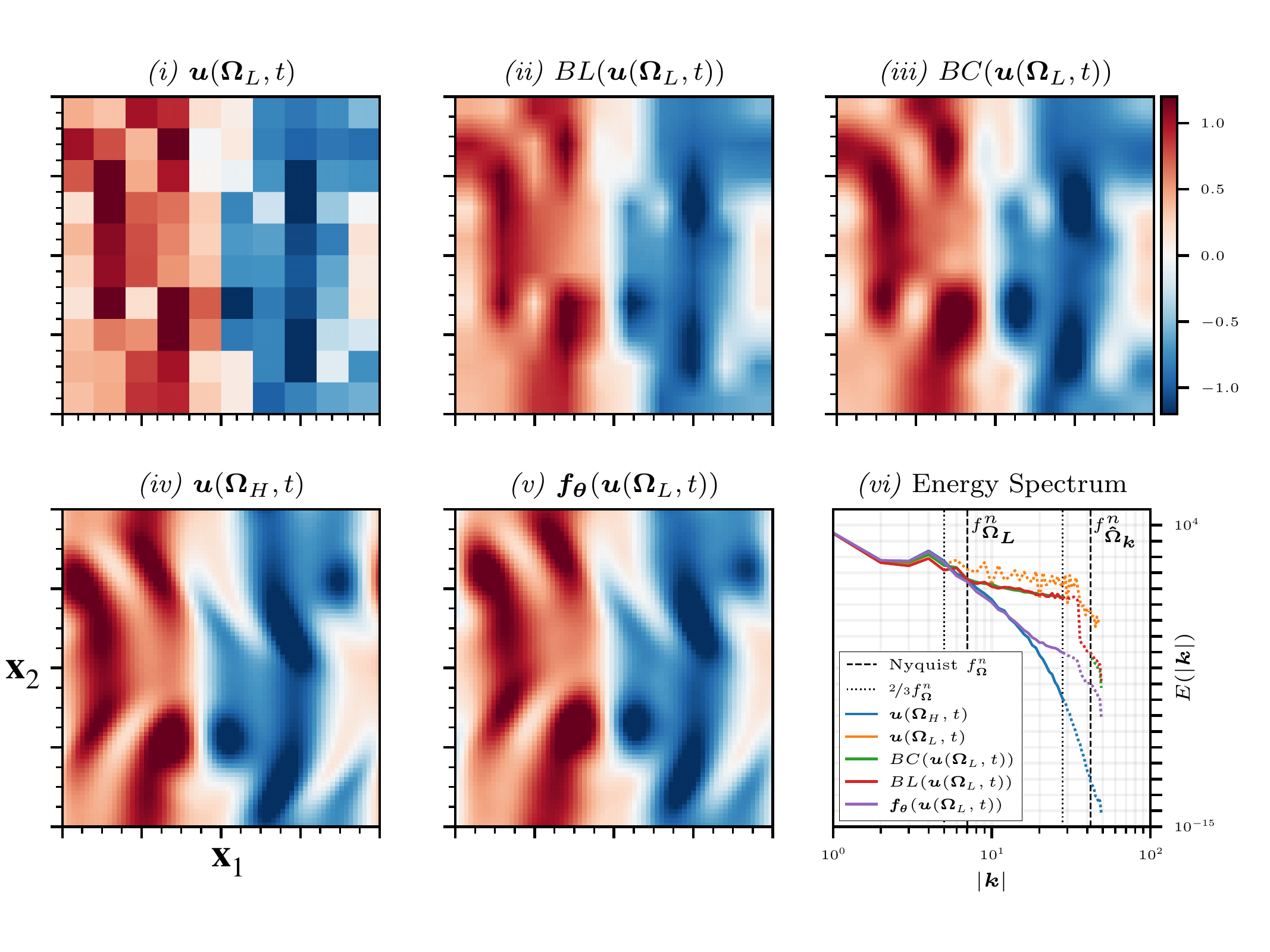}
    \caption{Reconstruction from sparse information: physics-constrained convolutional neural network (PC-CNN) compared with traditional interpolation methods to reconstruct a solution from a sparse grid ${\bm \Omega_L\in \mathbb{R}^{10 \times 10}}$ (100 points) to a high-resolution grid ${\bm \Omega_H\in \mathbb{R}^{70 \times 70}}$ (4900 points). Panel
        \textit{(i)}~shows the low-resolution input, ${\bm u}({\bm \Omega_L}, t)$;
        \textit{(ii)}~bi-linear interpolation, $\textit{BL}({\bm u}({\bm \Omega_L}, t))$;
        \textit{(iii)}~bi-cubic interpolation, $\textit{BC}({\bm u}({\bm \Omega_L}, t))$; 
        \textit{(iv)}~true high-resolution field, ${\bm u}({\bm \Omega_H}, t)$; 
        \textit{(v)}~model prediction of the high-resolution field, ${\bm {f_\theta}}({\bm u}({\bm \Omega_L}, t))$; 
        and \textit{(vi)}~energy spectra for each of the predictions. 
    Vertical lines $f^n$ denote the Nyquist frequencies for spectral grid, $f^n_{\bm {\hat{\Omega}_k}}$, and for the low-resolution grid, $f^n_{\bm {\Omega_L}}$.
    }
    \label{fig:interpolation}
\end{figure*}

Figure~\ref{fig:interpolation} shows a snapshot of results for the streamwise component of the velocity field, comparing network realisations ${\bm {f_\theta}}({\bm u}({\bm \Omega_L}, t))$ with the interpolated alternatives. Bi-linear and bi-cubic interpolation are denoted by $BL({\bm u}({\bm \Omega_L}, t)), BC({\bm u}({\bm \Omega_L}, t))$ respectively. We observe that network realisations of the high-resolution solution yield qualitatively more accurate results as compared with the interpolation. Artefacts indicative of the interpolation scheme used are present in both of the interpolated fields, whereas the network realisation captures the structures present in the high-resolution field correctly. Across the entire solution domain the model ${\bm {f_\theta}}$ achieves a relative error of $e=6.972 \times 10^{-2}$ compared with $e=2.091 \times 10^{-1}$ for bi-linear interpolation and $e=1.717 \times 10^{-1}$ for bi-cubic interpolation.

Although the relative error provides a notion of predictive accuracy, it is crucial to assess the physical characteristics of the reconstructed field~\citep{pope2000turbulent}. The energy spectrum, which is characteristic of turbulent flows, represents a multi-scale phenomenon where energy content decreases with the wavenumber. From the energy spectrum of the network's prediction, ${\bm {f_\theta}}({\bm u}(\bm{\Omega}_L, t))$, we gain physical insight into the multi-scale nature of the solution. Results in Figure~\ref{fig:interpolation} show that the energy content of the low-resolution field diverges from that of the high-resolution field, which is a consequence of spectral aliasing. Network realisations ${\bm {f_\theta}}({\bm u}({\bm \Omega_L}, t))$ are capable of capturing finer scales of turbulence compared to both interpolation approaches, prior to diverging from the true spectrum as $\lvert {\bm k} \rvert = 18$. These results show the efficacy of the method, recovering in excess of $99\%$ of the system's total energy. The physics loss, $\mathcal{L}_{P}$, enables the network to act beyond simple interpolation. The network is capable of de-aliasing, thereby inferring unresolved physics. Parametric studies show similar results across a range of scale factors $\kappa$ (shown in~\ref{app:kappa_results}).

\FloatBarrier

\section{Conclusions} \label{sec:conclusion}
We propose a physics-constrained convolutional neural network (PC-CNN) to solve inverse problems that are relevant to spatiotemporal partial differential equations (PDEs). 
First, we introduce the physics-constrained convolutional neural network (PC-CNN), which provides the means to compute the physical residual, i.e. the areas of the field in which the physical laws are violated by the prediction provided by the convolutional neural network (CNN).  
We formulate an optimisation problem by leveraging prior knowledge of the underlying physical system to regularise the predictions from the PC-CNN. This is augmented by a simple time-windowing approach for the computation of temporal derivative. 
Second, we apply the PC-CNN to solve an inverse problem in which we are given data that is contaminated by a spatially varying systematic error (i.e., bias, or epistemic uncertainty). The task is to uncover from the biased data the true state, which is the solution of the PDE. We numerically test the method to remove biases from data generated by three partial differential equations of increasing complexity (linear and nonlinear diffusion-convection, and Navier-Stokes). The PC-CNN successfully infers the true state, and the bias as a by-product, for a large range magnitudes and degrees of non-convexity. This demonstrates that the method is accurate and robust. 
Third, we apply the PC-CNN to solve an inverse problem in which we are given sparse information on the solution of a PDE. The task is to reconstruct the solution in space with high-resolution but without using the full high-resolution data in the training. We reconstruct the spatiotemporal chaotic solution of the Navier-Stokes equations on a high-resolution grid from a sparse grid containing only $\lesssim 1\%$ of the information. We demonstrate that the proposed PC-CNN provides accurate physical results, both qualitatively and quantitatively, as compared to traditional interpolation methods, such as bi-cubic interpolation. 
For both inverse tasks, we investigate the physical consistency of the inferred solutions for the two-dimensional turbulent flow. We find that the PC-CNN predictions provide correct physical properties of the underlying partial differential equation (Navier-Stokes), such as energy spectrum.  
This work opens opportunities for solving inverse problems with nonlinear partial differential equations. Future work will be focused on experimental data.

\paragraph{Acknowledgments}
The authors are grateful to G. Rigas for insightful discussions on turbulence.

\paragraph{Funding Statement}
D.Kelshaw. \& L.Magri. acknowledge support from the UK Engineering \& Physical Sciences Research Council, and ERC Starting Grant PhyCo 949388. L.M. acknowledges support from the grant EU-PNRR YoungResearcher TWIN ERC-PI\_0000005.

\paragraph{Data Availability Statement}
Codes and data are accessible via GitHub: \url{https://github.com/MagriLab/pisr}; \url{https://github.com/MagriLab/picr}; and \url{https://github.com/MagriLab/kolsol}.

\bibliographystyle{apalike}
\bibliography{references.bib}

\clearpage

\begin{appendix}
\section{Architectural Details}

Here, we provide details for the network architectures used to parameterise the mappings employed for each of the inverse problems.

\subsection{Uncovering solutions from biased data} \label{app:network:picr}

\begin{figure*}[!htb]
    \centering
    \includegraphics[width=\linewidth]{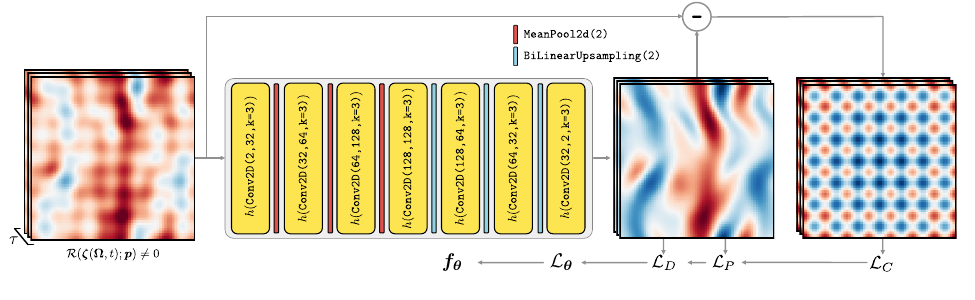}
    \cprotect
    \caption{Uncovering solutions from biased data. The model ${\bm {\eta_\theta}}$ is responsible for mapping the biased state ${\bm \zeta}({\bm \Omega}, t)$ to the true solution ${\bm u}({\bm \Omega}, t)$. Convolutional layers are parameterised as \verb|torch.Conv2D(c_{in}, c_{out}, k)|, where \verb|c_{in}, c_{out}| denote the number of input and output channels, respectively; and \verb|k| is the spatial extent of the filter. The term $h$ represents the activation, \verb|tanh| in this case. The terms $\mathcal{L}_D, \mathcal{L}_P, \mathcal{L}_C$ denote the data loss, physics loss, and constraint loss respectively. The combination of these losses forms the objective loss $\mathcal{L}_{\bm \theta}$, which is used to update the network's parameters. The term $\tau$ denotes the number of contiguous time-steps passed to the network, required for computing temporal derivatives. We provide an in-depth explanation of these losses in Sec.~\ref{sec:loss_terms}.}
    \label{fig:unet_architecture}
\end{figure*}

For the bias removal task, we use a UNet architecture to extract relevant features from the biased field. The architecture, as well as the general training loop can be seen in Figure~\ref{fig:unet_architecture}. This particular network configuration was found empirically. The input and output layers each feature two channels, corresponding to the two velocity components of the flow being studied. The network does not overfit, also thanks to the physics-based regularisation term in the loss function.

\subsection{Reconstruction from sparse information} \label{app:network:pisr}

\begin{figure*}[!htb]
    \centering
    \includegraphics[width=\linewidth]{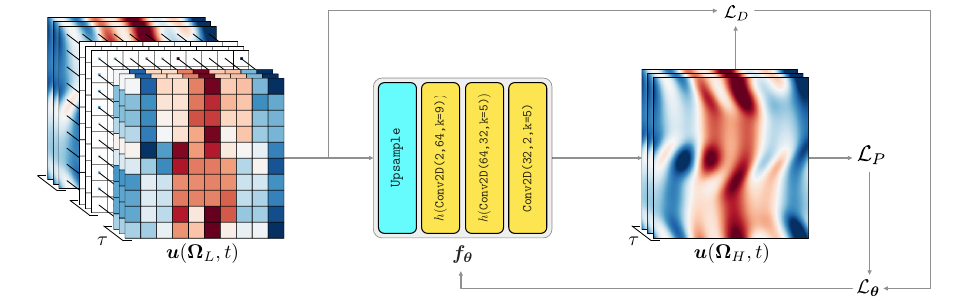}
    \cprotect
    \caption{Reconstruction from sparse information. The model ${\bm {f_\theta}}$ is responsible for mapping the low-resolution field ${\bm u}({\bm \Omega_L}, t)$ to the high-resolution field ${\bm u}({\bm \Omega_H}, t)$. The upsampling layer (blue) performs bi-cubic upsampling to obtain the correct spatial dimensions. Convolutional layers are parameterised as \verb|torch.Conv2D(c_{in}, c_{out}, k)|, where \verb|c_{in}, c_{out}| denote the number of input and output channels, respectively; and \verb|k| is the spatial extent of the filter. The term $h$ represents the activation, \verb|tanh| in this case. The terms $\mathcal{L}_D, \mathcal{L}_P$ denote the data loss and physics loss, respectively, the combination of which forms the objective loss $\mathcal{L}_{\bm \theta}$, which is used to update the network's parameters, ${\bm \theta}$. The term $\tau$ denotes the number of contiguous time-steps passed to the network, required for computing temporal derivatives. We provide an in-depth explanation of these losses in Sec.~\ref{sec:loss_terms}.}
    \label{fig:reconstruction_model}
\end{figure*}

For the reconstruction task, a three-layer convolutional neural network is used for the model. Figure~\ref{fig:reconstruction_model} provides an overview of the architecture, as well as the general training loop. The baseline network configuration is similar to that of~\citet{Dong2014}, with the additional constraints imposed through the physics-based loss function. The input and output layers each feature two channels, corresponding to the two velocity components of the flow being studied.

\section{Pseudo-spectral discretisation of Navier-Stokes residual} \label{app:pseudospectral_discretisation}
Operating in the Fourier domain eliminates the diverge-free constraint \citep{canuto1988SpectralMethodsFluid}. The equations for the spectral representation of the Kolmogorov flow are
\begin{equation}
    \mathcal{R}(\bm{\hat{\bm u}}_k; {\bm p}) = \left( \tfrac{d}{dt} + \nu \lvert {\bm k} \rvert^2 \right) \bm{\hat{\bm u}}_k - {\bm {\hat f}}_k + {\bm k} \frac{{\bm k} \cdot {\bm {\hat f}}_k}{\lvert {\bm k} \rvert^2} - {\hat {\bm g}}_k,
\end{equation}
with ${\bm {\hat f}}_k = - \left( \widehat{{\bm u} \cdot \nabla {\bm u}} \right)_{k}$, where nonlinear terms are handled pseudospectrally, employing the $\nicefrac{2}{3}$ de-aliasing rule to avoid unphysical culmination of energy at the high frequencies~\cite{canuto1988SpectralMethodsFluid}.
A solution is produced by time-integration of the dynamical system with the explicit forward-Euler scheme, choosing a time-step $\Delta t$ that satisfies the Courant-Friedrichs-Lewy (CFL) condition. Initial conditions are generated by producing a random field scaled by the wavenumber, which retains the spatial structures of varying lengthscale in the physical domain~\citep{ruan1998}.  

\section{Multi-wave number bias removal} \label{app:multi_freq}
\begin{figure*}[!htb]
    \centering
    \includegraphics[width=0.95\linewidth]{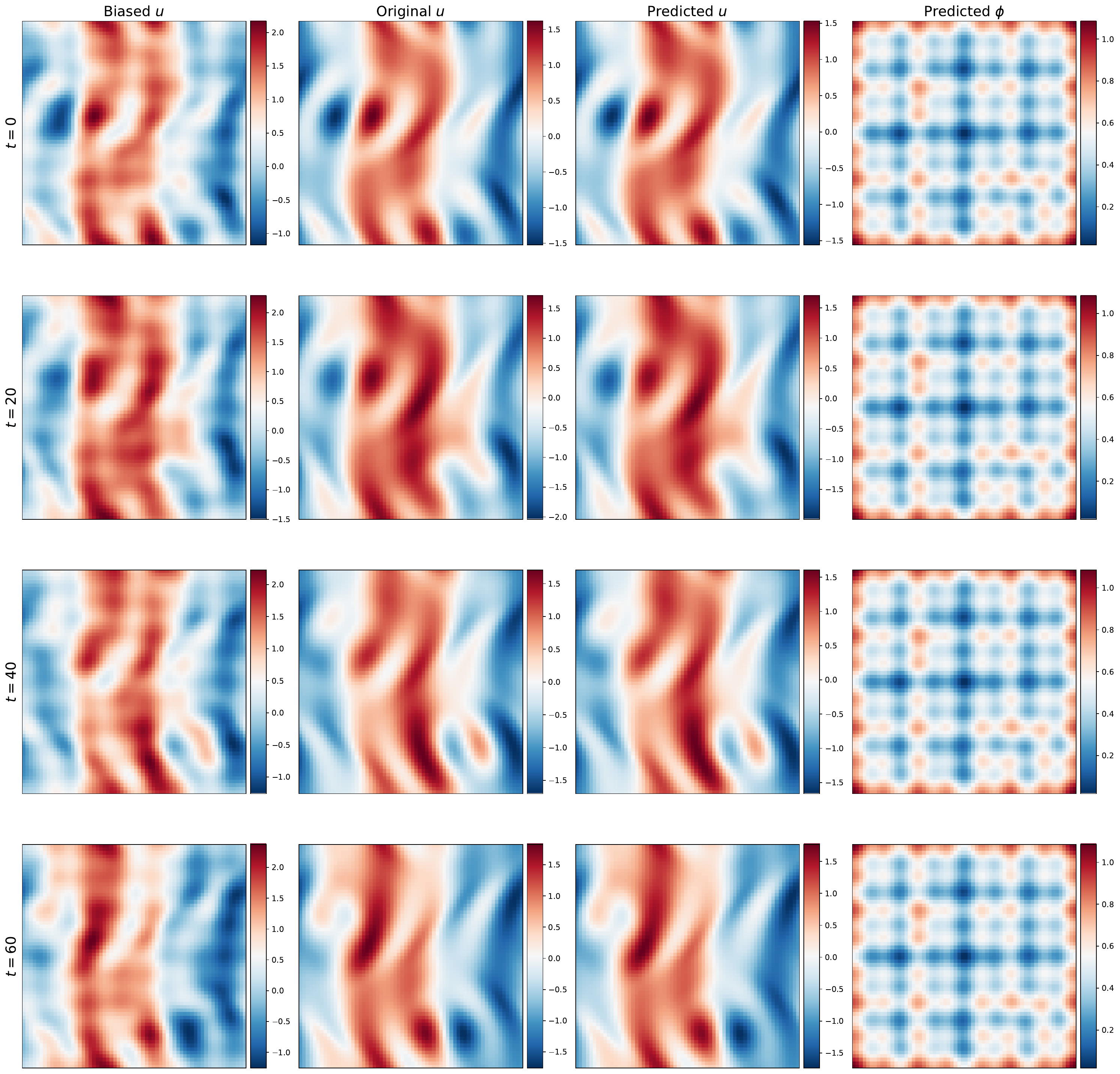}
    \cprotect
    \caption{Removal of a stationary bias field containing multiple wave numbers. The first three columns show the biased, original, and predicted flow fields respectively. The rightmost column depicts the predicted additive bias. Results are shown for multiple time-steps in the simulation trajectory.}
    \label{fig:app:multi_freq}
\end{figure*}

Further to the parameterisation outlined in Section~\ref{sec:bias_parameterisation}, we consider stationary biases that contain multiple wave numbers. We define this composite bias as the mean of two bias fields, with $k_\phi \in \{3, 7\}, \mathcal{M} = 0.5$. Bias removal was performed for the Kolmogorov flow case, following the same procedure as outlined in Section~\ref{sec:bias_removal_methodology}. Results for the bias removal achieved a relative error of $5.83 \times 10^{-2}$, with results for multiple time-steps shown in Figure~\ref{fig:app:multi_freq}. The structure of the predicted bias captures the extrema of the bias, however there is small discrepancy in the magnitude as the field should be symmetric along both axes. This further demonstrates the robustness of the proposed approach.

\section{Parametric study on the scale factor} \label{app:kappa_results}

\begin{figure*}[!htb]
    \centering
    \includegraphics[width=\linewidth]{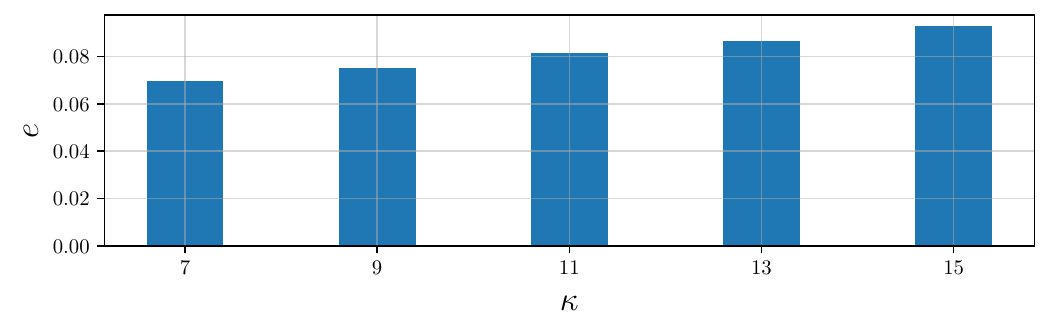}
    \cprotect
    \caption{Reconstruction from sparse information. Performance for the reconstruction task are shown for range of scale factors $\kappa$, measuring performance with the relative $\ell^2$-errors, $e$, between the predicted high-resolution fields ${\bm {f_\theta}}({\bm u}({\bm {\Omega_L}}, t))$, and the true high-resolution fields ${\bm u}({\bm {\Omega_H}}, t)$.}
    \label{fig:sr_sweep}
\end{figure*}

Experiments for reconstruction from sparse information were also run for $\kappa \in \{7, 9, 11, 13, 15\}$ using the same experimental setup; ultimately, demonstrating reconstruction results for a range of scale factors. Results for each of these scale factors are shown in Figure~\ref{fig:sr_sweep}. We observe that the relative $\ell^2$-error increases with increasing $\kappa$; this is to be expected as the available information becomes more sparse.

\end{appendix}
\end{document}